\documentclass[10pt,journal,compsoc]{IEEEtran}

\usepackage{balance}
\usepackage{graphicx}
\usepackage{amssymb}
\usepackage{verbatim}
\usepackage[hyphens]{url}
\usepackage{hyperref}
\usepackage{tikz}
\usepackage{array}
\usepackage{pgfplots}
\pgfplotsset{compat=1.14}
\usepackage{graphicx}
\usepackage{subfig}
\usetikzlibrary{arrows,shapes,positioning,shadows,trees}

\usepackage{forest}
\usetikzlibrary{arrows.meta}

\tikzstyle{arrow} = [thick,->,>=stealth]

\begin{document}

\title{Taxonomy of Big Data: A Survey}

\author{Ripon Patgiri\\National Institute of Technology Silchar %
\IEEEcompsocitemizethanks{\IEEEcompsocthanksitem Ripon Patgiri, Sabuzima Nayak, and Samir Kumar Borgohain, Department of Computer Science \& Engineering, National Institute of Technology Silchar, Assam, India-788010 \protect\\
E-mail: ripon@cse.nits.ac.in}
}

\markboth{}%
{Patgiri \MakeLowercase{\textit{et al.}}: Taxonomy of Big Data}

\IEEEtitleabstractindextext{%
\begin{abstract}
The Big Data is the most popular paradigm nowadays and it has almost no untouched area. For instance, science, engineering, economics, business, social science, and government. The Big Data are used to boost up the organization performance using massive amount of dataset. The Data are assets of the organization, and these data gives revenue to the organizations. Therefore, the Big Data is spawning everywhere to enhance the organizations' revenue. Thus, many new technologies emerging based on Big Data. In this paper, we present the taxonomy of Big Data. Besides, we present in-depth insight on the Big Data paradigm.
\end{abstract}

}

\maketitle

\IEEEdisplaynontitleabstractindextext

\IEEEpeerreviewmaketitle

\ifCLASSOPTIONcompsoc
\IEEEraisesectionheading{\section{Introduction}\label{sec:introduction}}
\else
\section{Introduction}
\label{sec:introduction}
\fi
\IEEEPARstart{Big}{Data} Big Data growing rapidly day-by-day due to producing a huge data from various sources. The new emerging technologies act as catalyst in growing data where the growth of data get an exponential pace. For instance, IoT. Moreover, smart devices engender enormous data. The smart devices are core component of smart cities (includes smart healthcare, smart home, smart irrigation, smart schools and smart grid), smart agriculture, smart pollution control, smart car and smart transportation \cite{A8,Lin}. Data are generated not only by IoT devices, but also sciences, business, economics and government. The science generates humongous dataset and these data are handled by Big Data. For example, Large Hadron Collider in Geneva. Moreover, the web of things also a big factor in engendering the huge size of data. In addition, the particle analysis requires a huge data to be analyzed.  Moreover, the seismology also generates large dataset, and thus, the Big Data tools are deployed to analyze and predict. Interestingly, the Big Data tools are deployed in diverse fields to handle very large scale dataset. There are hundreds of application field of the Big Data which makes the Big Data paradigm glorious in this high competitive era. 

The paper present following key point-
\begin{itemize}
\item Provides rich taxonomy of Big Data.
\item Presents Big Data in every aspect precisely.
\item Highlights on each technology.
\end{itemize}

\begin{figure}
\centering
\newcolumntype{C}[1]{>{\centering}p{#1}}
\begin{forest}
  	for tree={
  		if level=0{align=center}{
    		align={@{}C{40mm}@{}},
  		},
  		grow=east,
  		draw,
  		font=\sffamily\bfseries,
  		edge path={
    		\noexpand\path [draw, \forestoption{edge}] (!u.parent anchor) -- +(4mm,0) |- (.child anchor)\forestoption{edge label};
  		},
  		parent anchor=east,
  		child anchor=west,
  		l sep=10mm
  }
  [Big Data
  	[Security and Privacy]
  	[Data Mining \& Machine Learning]
    [Big Data Management]
    [Storage System]
    [Compute Infrastructure]
  	[Semantic]
  ]
\end{forest}
\caption{Taxonomy of Big Data Technology}
\label{taxo}
\end{figure}
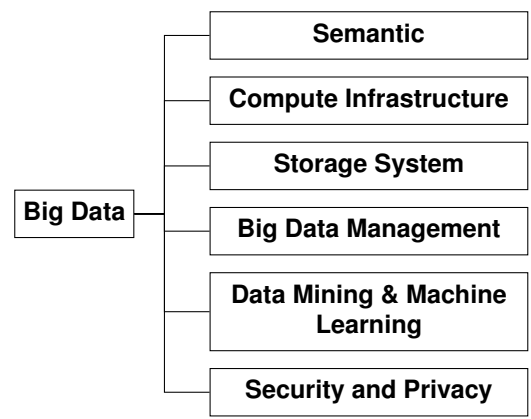

The Big Data is categorized into seven key categories, namely, semantic, compute infrastructure, storage system, Big Data Management, Big Data Mining, Big Machine Learning, and Security \& Privacy as shown in the figure ~\ref{taxo}. The paper discusses on semantic on the Big Data and explore $V_3^{11}+C$ \cite{RP} in the section ~\ref{sem}. The paper also discusses on compute infrastructure and classifies the Big Data into three categories, namely, MapReduce, Bulk Synchronous Parallel, and Streaming in the section ~\ref{com}. Besides, the Big Data storage system is classified into four categories, namely, storage architecture, storage implementation, storage structure and storage devices in section ~\ref{ss}.

\section{Semantic of Big Data}
\label{sem}

\begin{figure*}
\centering
\includegraphics[width=0.8\textwidth]{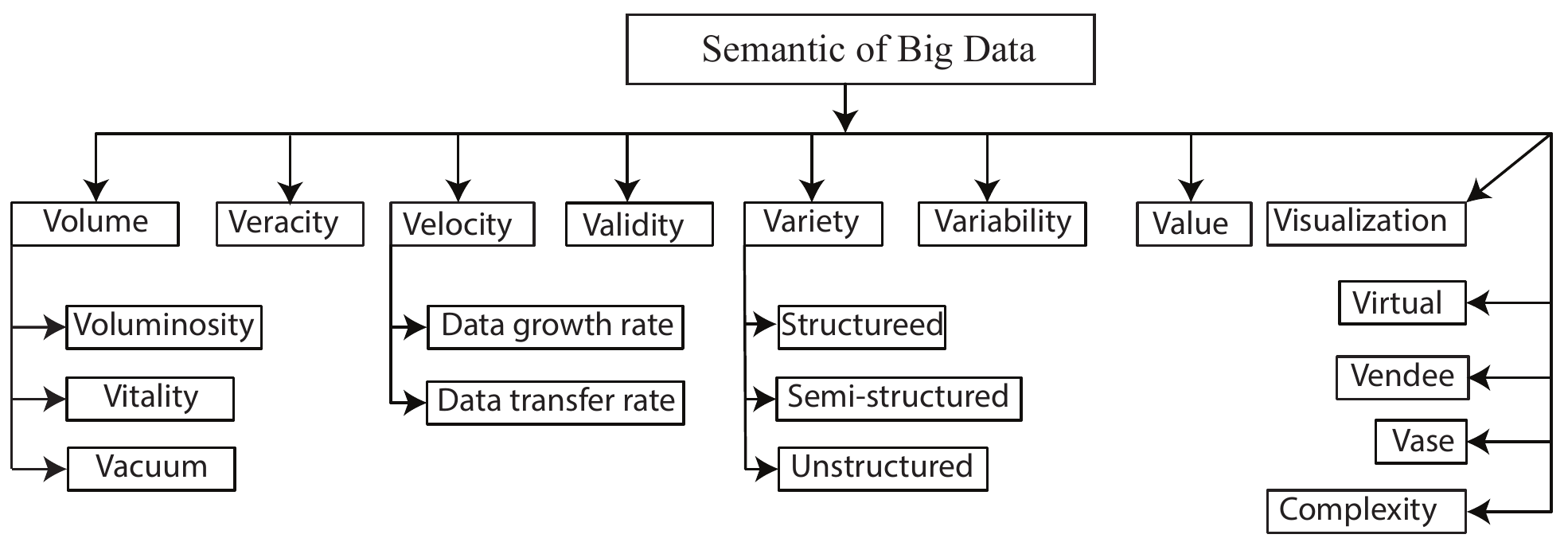}
\caption{Semantic of Big Data. Source \cite{RP}}
\label{Vs}
\end{figure*}

There are many V's coming up to define the characteristics of Big Data. Doug Laney defines Big Data using 3V's, namely, volume, velocity and variety. Now, the $V_3^{11}+C$ is used to define the characteristics of Big Data \cite{RP}. The different kind of V's are shown in the figure ~\ref{Vs}. The volatility and visibility is not the family of V \cite{RP}. The table ~\ref{tab} defines the meaning of Vs precisely.

\begin{table*}[ht]
\centering
\caption{Individual meaning the V family}
\begin{tabular}{p{2cm}p{4cm}p{6.5cm}}
\hline
\textbf{Name} & \textbf{Short meaning} & \textbf{Big Data context}\\ \hline
Volume	& Size of data & Voluminosity, Vacuum and Vitality\\
Velocity & Speed & Transfer rate \& Growth rate of data\\
Variety & Numerous types of data & Structured, unstructured and semi-structured\\
Veracity & Accuracy and truthfulness & Accuracy of Data\\
Validity & Cogency & Correct Data \\
Value & Worth & Giving worth to the raw data\\
Virtual & Nearly actual & Managing large number of data.\\
Visualization & To be shown & Logically display the large-set of data.\\
Variability & Change & Change due to time and intention\\
Vendee & Client management & Client management and fulfilling the client requirements\\
Vase & Big Data Foundation & IoT, Cloud Computing etc.\\
Complexity & Time and Space requirement & Computational performance\\
\hline
\end{tabular}
\label{tab}
\end{table*}

\section{Compute Infrastructure}
\label{com}

\begin{figure}[ht]
\centering
\includegraphics[width=0.45\textwidth]{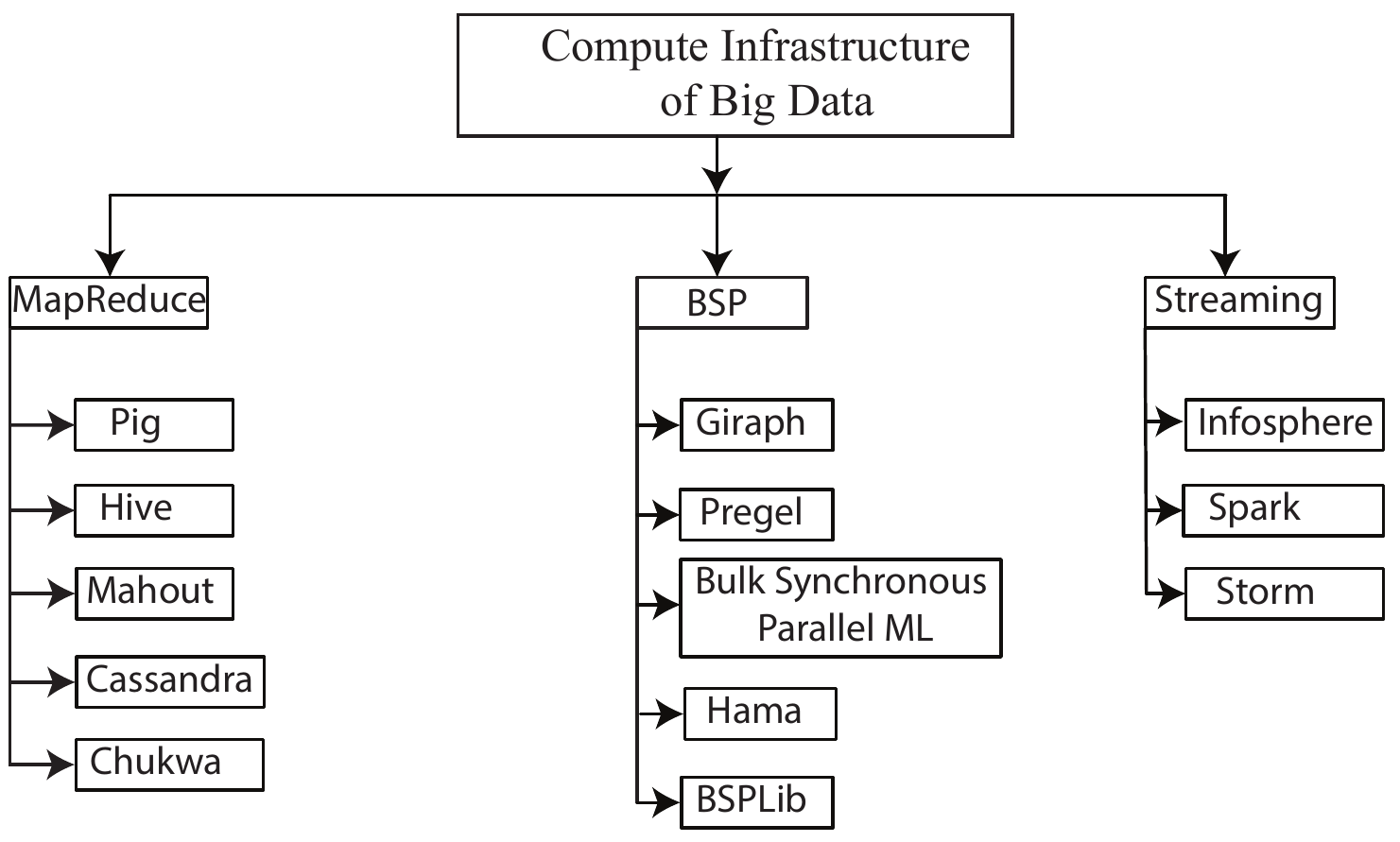}
\caption{Compute Infrastructures}
\label{comp}
\end{figure}

\begin{figure*}[ht]
\centering
\includegraphics[width=0.7\textwidth]{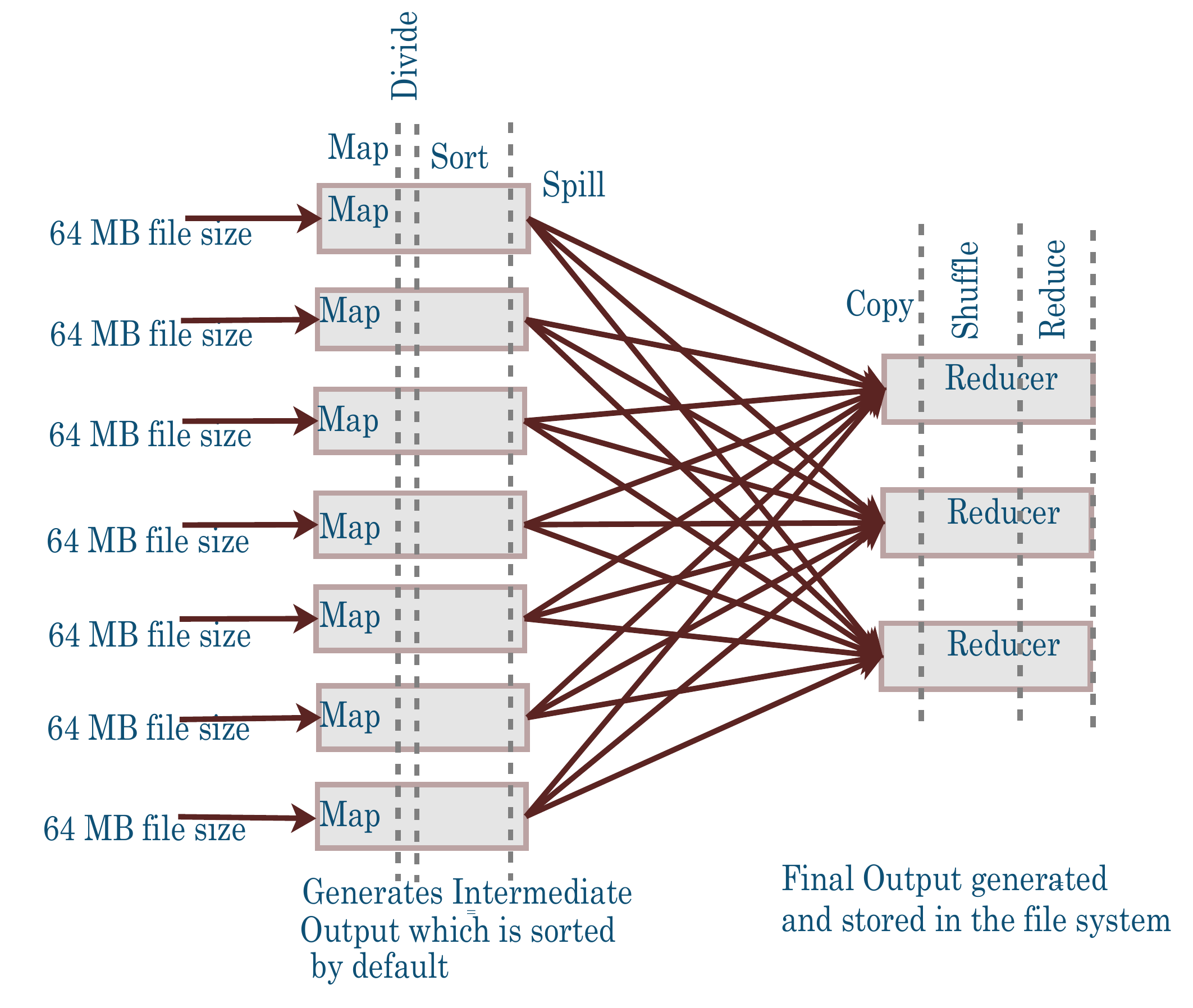}
\caption{Architecture of MapReduce}
\end{figure*}

\subsection{MapReduce}
\label{prog}
The MapReduce \cite{MR} programming paradigm is the best parallel programming concept even if the MapReduce is purely based on the only Map and Reduce task. However, the DryadLINQ \cite{DryadLINQ} is also emerging based on Microsoft Dryad \cite{Dryad}. But, MapReduce can solve almost every problem of distributed and parallel computing, and large-scale data-intensive computing. It has been widely accepted and demonstrated that the MapReduce is the most powerful programming for large scale cluster computing. The conventional DBMS is designed to work with structured data and it can scale with scaling of expensive hardware, but not low-cost commodity hardware. The MapReduce programming works on low-cost unreliable commodity hardware, and it is an extremely scalable RAIN cluster, fault tolerant yet easy to administer, and highly parallel yet abstracted \cite{Sherif13}. The MapReduce is key/value pair computing wherein the input and output are in the form of key and value. 
\begin{description}
    \item[Map.] The Map function transforms the input key/value pair to intermediate key-value pair \cite{MR}. For instance, the key is the file name and value is its content. The output of Map function is transferred to reducer function by shuffling and the sorting. Sorting is done by some internal sorting techniques (Timsort for internal sorting or quick sort). The input is fetched from the file system, namely, HDFS \cite{HDFS}, and GFS \cite{GFS}. The input size varies from 64 MB to 1GB and the performance can vary in a specific range, which is evaluated in the paper \cite{Dev15}.
    \item[Reduce.] The Reduce function consists of three different phases, namely, copy phase, shuffle phase and reduce phase. The copy phase fetches the output from the Map function \cite{Late}. The Map function spills the output when it reaches a specific size (for example, 10MB). This spilling cause early starting of Reduce task, otherwise the reduce task has to wait until the Map task has not finished \cite{Condie10}. The shuffle phase sort the data using an external sorting algorithm \cite{Late}. The shuffle phase and copy phase takes a longer time to execute as compared other phases. The reduce phase computes as the user defines and produces final output \cite{Late}. The final output is written to the file system.
\end{description}

 The limitation of MapReduce is very limited and these are listed below:
 \begin{enumerate}
    \item The computation depends on the previously computed value is not suitable for MapReduce.
    \item If the computation cannot be converted into the Map and Reduce form.
    \item MapReduce is not suitable for small-scale computing, such like RDBMS jobs.
    \item No high-level language \cite{Lee11}: The MapReduce does not support SQL. 
    \item No schema and no index \cite{Lee11}: The MapReduce does not support schema and index, since, MapReduce is a programming language.
    \item A Single fixed data flow \cite{Lee11}: The MapReduce strictly follows key/value computation style, and Map and Reduce function. 
\end{enumerate}

Withal, the SQL can be converted to MapReduce programming and can easily be made using these two roles (map and reduce). Almost all distributed system's problems can be solved by MapReduce, albeit, folk can claim that the conversion of their task to the Map and Reduce functions is difficult and performing low. Antithetically, the paper \cite{Christos14} reveals MapReduce as the query processing engine. Moreover, the MapReduce can be used for indexing, schema support, structured, unstructured, and semi-structured information. MapReduce is also used to process Big Graph \cite{DKS}. It depends on the programming way and fine tune of the programmer. Hence, the MapReduce is the most powerful engine to work out any sort of distributed system.

\subsubsection{Achilles' Heels}
The MapReduce consists of Map and Reduce task spawned in several machines for maximizing the parallelism. The tasks are split into several workers and assigned them to process. Suppose, one of the workers become straggler \cite{Rajdeep,Deep}, then the entire process becomes Achilles' Heels. That is, one task can complete a job $J$ in 10 minutes, then 10 tasks should complete the job $J$ in one Minutes. Unfortunately, one task is taking 10 minutes, then the time to complete the job is 10 minutes, even though the task are running in parallel and works are divided equally likely.
 
Almost every problem is solvable in MapReduce, but, the solution may not be suitable or may not perform well. A deliberate design and optimization are required for MapReduce to do comfortably. For instance, thousands of Map with one Reduce task is obviously slower than many reduce tasks with thousands of Map tasks. Some investigation on the unsuitability of conventional MapReduce is given below:
    \begin{enumerate}
        \item The Reducer Task takes more time to finish. If the reducer task becomes slow or straggler, then there is no guarantee that the job will be completed same time period with straggler and without straggler. The copy of a straggler task can be scheduled in another suitable node which increases network traffic. Making the decision of whether reduce task has to be rescheduled or not is a challenge itself. For instance, reduce task is about to complete and straggling. In this situation, the scheduler must examine whether scheduling a copy of that straggler task is profitable or not. The SkewTune \cite{Kwon12} schedules a partial copy of straggler task, where the partial copy is the remaining task to be finished.
        \item The MapReduce does redundant computing in some situation \cite{Kalavri13}. For example, MapReduce has completed a job, word count problem. Now, there is one or few word changes in the entire file system. The MapReduce recompute the entire word count job and it does not reuse the previous result of a computation of the same job. This problem has been addressed by ReStore \cite{Elghandour12}, and  Early Accurate Result Library (EARL) \cite{Laptev12}.
        \item The MapReduce shows certain disadvantage in reading input several times redundantly, while the same job is run several times on the same data.
        \item A careful implementation of MapReduce has to be done, while the data are coming continuously and writing to the file system. In this continuous online data degrades the performance of MapReduce.
        \item The sharing among multiple jobs can improve the performances \cite{Kalavri13}. This challenge is addressed in MRShare \cite{Nykiel10}.

    \end{enumerate}

\subsection{BSP}

\subsubsection{Giraph}
Giraph \cite{Giraph} is an open source system which is used for graph processing on big data. It uses MapReduce implementation for graph processing. In general, it follows a master/workers model. Also it support multithreading by assigning each worker multiple graph partitioning. During each superstep, an available worker pairs compute threads with uncomputable partitions. And between supersteps, workers perform serial tasks (e.g. blocking on global barriers, resolving mutations) by executing with a single thread. Moreover Giraph implement BSP by maintaining two message stores for holding messages from previous and current supersteps respectively. It reduces memory usage and computation time by using receiver-side message combining. Also for global coordination or counters blocking, aggregators are used. And  master.compute() is used for serial computations at the master.

\subsubsection{Pregel}
Pregel \cite{Pregel} is a system that provides a graph processing API along with BSP \cite{Grzegorz} with a vertex-centric, programming model. Its programs are inspired by Valiant’s Bulk Synchronous Parallel model \cite{Valiant}. Directed graph is input to a Pregel computation where each vertex have a string vertex identifier for unique identification. A typical Pregel computation have input (graph is initialized), then it have sequence of supersteps separated by global synchronization points, finally it give the output and then algorithm terminates. In each superstep the vertices compute in parallel while executing the same user-defined function expressing the logic of the given algorithm. However, algorithm terminates when every vertex vote to halt. Pregel also have aggregators. Aggregator is a mechanism for global data, monitoring, and communication. And it can be used for global coordination

To improve usability and performance Pregel keeps vertices and edges on the machine doing computation. And it uses network only for messages. Also Pregel programs are deadlock free. Moreover algorithms developed by Pregel can be used to solve real problems such as Shortest Paths, Page Rank, Bipartite Matching, Semi-Clustering algorithm and so on.

\subsubsection{BSP ML} 
Bulk Synchronous Parallel ML language (BSML) \cite{Loul} is a library for parallel programming along with functional language Objective Caml. It is based on an extension of the λ-calculus by parallel operations on parallel vector. Parallel vector is a parallel data structure. Moreover, the BSML library provide a safe environment for declarative parallel programming. And programs are similar to functional programs (in Objective Caml) but using few additional functions. Furthermore, BSML have provided functions. These functions are used for accessing the parameters of the parallel machine for creating and performing operations on parallel vectors. 

BSML is based on a confluent extension of the λ-calculus, making it deterministic and deadlock free. In BSML, programs can be easily composed, written and reused. Also it has simpler semantics and better readability. And it is implemented using Objective Caml in a modular approach.This approach helps to communicate with various communication libraries, makes it portable, and efficient, on a wide range of architectures.

\subsubsection{Hama}
Hama \cite{Hama} is a pure bulk synchronous parallel model which can do vast scientific computations, e.g. matrix, graph, and network algorithms. It’s internal architecture is different from other known computational frameworks because of its underlying BSP based communication and synchronization mechanisms. Again it is based on Master-Slave model consisting of three major components, BSP master, Groom Server, and Zookeeper. Some functions of BSP master are scheduling jobs, task assignment to a Groom Server, maintainance of the Groom Server status and job progress information. Groom Server acts as a slave and it executes tasks assigned by the BSP Master. And Zookeeper gives efficient barrier synchronization to the BSP tasks.

The robust BSP model helps in avoiding deadlines and conflicts during communication in Hama. Also it is flexible so it can be used with any distributed file system. However BSP Master is a single point of failure and the application will stop if it dies. Additionaly the graph partitioning algorithm have to be customized, to avoid communication overhead between nodes. 

\subsubsection{BSPLib}
BSPLib \cite{BSPLib} is a small communication library for BSP programming. It is done in a Single Program Multiple Data (SPMD) manner. It consist of only 20 basic operations. It has two modes of communication, direct remote memory access (DRMA) and bulk synchronous message passing (BSMP) approach. BSPLib provides the infrastructure required for the user for data distribution, and communication required for changing parts of the data structure present in a remote process. Additionally, it provide a higher-level libraries or programming tools which is architecture independent and automatically distribute the problem domain among the processes. 

\subsection{Streaming}

\subsubsection{Infosphere}
InfoSphere \cite{InfoSphere} is a component-based distributed stream processing platform. The stream processing applications can be graphs of modular, reusable software components interconnected by data streams. Likewise, Component-based programming model allows composition and reconfiguration of individual components to create different applications. These applications can perform different types of analysis or answer different types of queries. Also it helps in creation and deployment of new applications without disturbing existing ones. It is used in sense-and-respond application domains. And it provides both language and runtime to these applications to improve efficiency in processing data from high rate streams. 

\subsubsection{Spark}
Spark \cite{Spark} is a new cluster computing framework. It supports applications with working sets while providing scalability and fault tolerance properties to MapReduce. Spark provides three data abstractions, resilient distributed datasets (RDDs), and two restricted types of shared variables: broadcast variables and accumulators. RDD represents a read-only collection of objects. These objects are partitioned across a set of systems and they can be rebuilt if a partition is lost. Suppose a large read-only piece of data (e.g., a lookup table) is used in multiple parallel operations, so it should be distributed to the workers only once. Similarly, Broadcast variable wraps the value and copy it once to each worker. Likewise, Accumulators are variables that workers use an associative operation to “add”, and these variables can only be read by driver.

\subsubsection{Storm}
Storm \cite{Storm1} software is a framework for building, processing applications that use the computing resources of all systems in a cluster. Based on varying processing needs of such applications, the platform should automatically grow and shrink as per requirement. Storm efficiently processes unbounded streams of data. It give the ability to users, to transform an existing stream into a new stream using two primitives,a spout and a bolt. A spout is a source of streams. Usually, user have to provide code that reads data from source (such as a queue, a database or a website). This data is then given to one or more bolts for processing. LIkewise, a bolt takes input streams, does processing, and then gives new streams of data. In addition, Storm cluster have two types of nodes, the master node and the processing nodes. The master node runs a daemon called Nimbus. Numbus is responsible for distribution of code around the cluster, task assignment, and monitoring their progression and failures. Whereas, Processing nodes run a daemon called Supervisor. Supervisor listens for work assignment to it's system. It starts and stops processes based on the work assigned by the master node.

\section{Storage System}
\label{ss}

\begin{figure}
\centering
\includegraphics[width=0.45\textwidth]{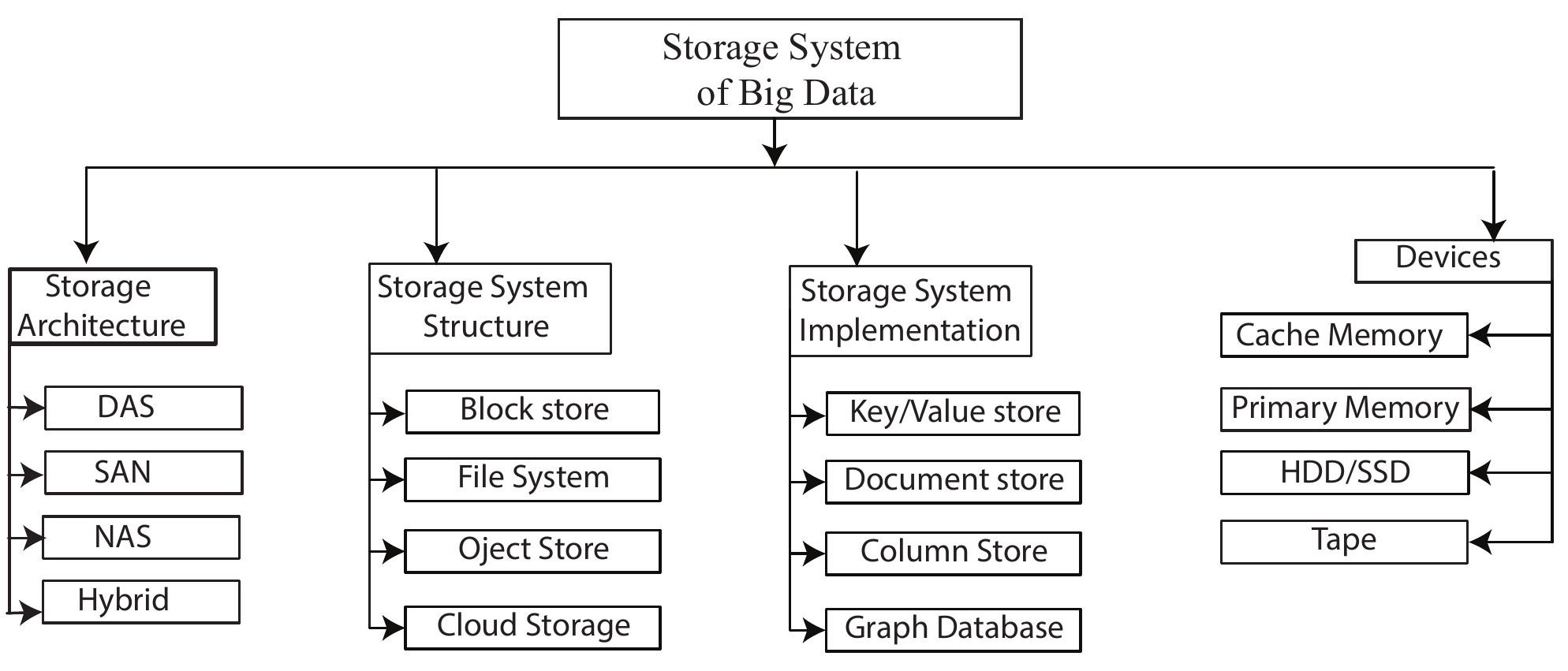}
\caption{Storage System}
\label{sst}
\end{figure}

\subsection{Storage Architecture}
\label{SA}
The storage system architecture is broadly categorized into three categories, namely, Direct-Attached Storage (DAS), Network Attached Storage (NAS), and Storage Area Network (SAN) \cite{Mesnier}. The three architectures have its own pros and cons, shown in table ~\ref{tab2}. 

\begin{table*}[ht]
\caption{Storage Architecture comparison of features}
\centering
\scalebox{0.8}{
\begin{tabular}{p{2cm}p{4cm}p{4cm}p{4cm}}
\hline
\textbf{Features} & \textbf{DAS} & \textbf{NAS} & \textbf{SAN} \\ \hline 
Data Transmission & IDE/SCSI & TCP/IP & Fibre \\ 
Storage Type & Track \& Sector & Shared Files & Blocks \\ 
Fault Tolerance & RAID & Replication & RAID \\ 
Scalability & Not Scalable & Scalable & Limited \\ 
Distance coverage & Within a system & Very Long Distance & Very Short Distance \\
Pros &  Very simple architecture, easy to manage, ideal for local services  & Unbound scalability, distance does not matter & Very fast access of storage device \\ 
Cons & Not scalable & Not fast as well as SAN &  Complex scalable, distance coverage is a problem. \\ \hline
\end{tabular}
}
\label{tab2}
\end{table*}

\subsubsection{Direct Attached Storage (DAS) }
DAS is digital storage, which attaches storage directly to the computer that accessing it \cite{Mesnier,Sacks}. These storages are from USB drive, and by Bus, i.e., every server has its own storage space directly attached to it without using network accessing.

\subsubsection{Network Attached Storage (NAS)}
The storage is attached through an Ethernet switch to scale the storage system. The NAS uses TCP/IP protocol to access the storage \cite{Mesnier,Sacks}. The application server is detached from the file system and data storage. The advantages of detaching application server, and file system \& storage system is incremental scalability. It is really easy to design a disaster recovery system using NAS. The performance is the major issue in the NAS.

\subsubsection{Storage Area Network (SAN)}
The storage devices are attached with fiber channel and storage are networked together \cite{Mesnier,Sacks}. Thus SAN strides the speed accessing of storage devices. Storage is connected through fiber switch so that the accessing the data become faster. The performance is the major advantage and scalability is the major issue in the SAN. The SAN supports fast accessing of data through a fibre channel. The SAN outperform NAS and DAS in performance, but NAS outperform SAN and DAS in scalability.

\subsection{Storage System Implementation}
A billion dollar question is how do we store the Exabytes of data? How do we process them efficiently? The answer is partially given by Apache Hadoop \cite{HDFS} and GFS \cite{GFS}. Another good example is Google Spanner \cite{Spanner} and Microsoft Dryad \cite{Dryad}. The assumed environment must not be Infiniband with a high-end server. No doubt, the production version is deployed in the high-end server configuration, but our assumption is a low-end server. To span the probable failure, the solution must assume, how to deploy thousands of low-cost commodity hardware. The low-cost commodity hardware is more vulnerable to failure, and therefore, the replication technique is used to overcome the failure rate and achieves maximum parallel processing and reliability of the system.

\begin{figure}[ht]
\centering
\includegraphics[width=0.45\textwidth]{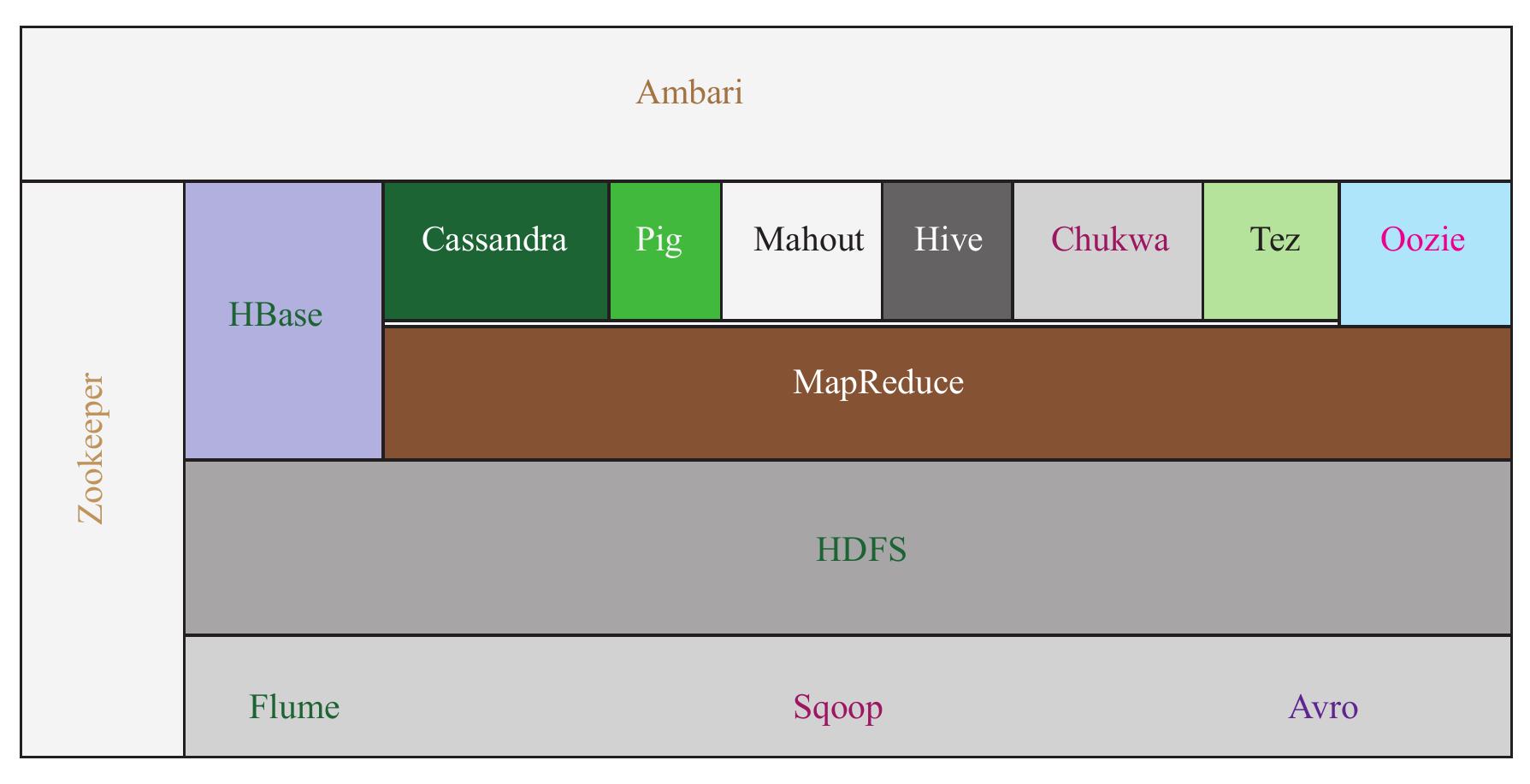}
\caption{The Hadoop Stack}
\label{hadoop}
\end{figure}

\begin{figure}[ht]
\centering
\includegraphics[width=0.45\textwidth]{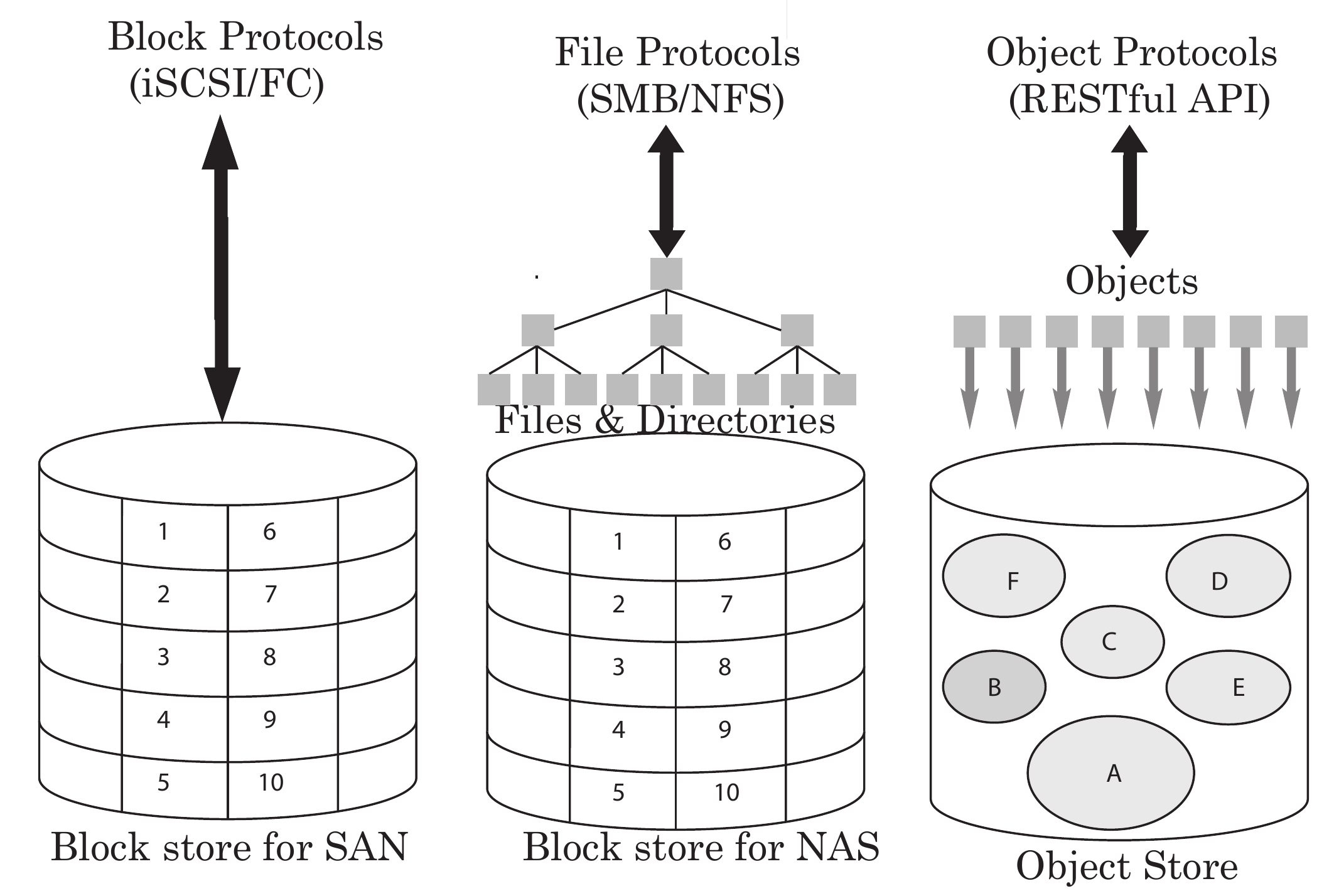}
\caption{Architecture of Block, File and Object store}
\label{ofb}
\end{figure}

\subsubsection{File System for Big Data}
The most popular file system like GFS \cite{GFS} and HDFS \cite{HDFS} need to enhance their scalability, fine-tune their performance in bigger dimension.  Conventionally, the file system and block storage are different, but both are combined to form a modern distributed file system as shown in figure ~\ref{ofb}. The file system holds some properties of a conventional file system and blocks storage system. The data are stored in the form of a file within a specified range of file size and split into block otherwise. Moreover, the files are kept same as originals if the file size falls within a specified range. A file size varies from MB to TBs, that involves some low sized files are mixed in concert to restrict from generating more number of Metadata and split high sized file to maximize parallelism. File system stores the data in a hard disk or SSD devices and store data information in RAM, known as Metadata \cite{NISO}, to enhance the access time. A dedicated Metadata server (MDS) serves client queries for data. The MDS become a bottleneck when smaller sized file. The billions of small sized file create no more storage space, but produce the huge size of Metadata, results performance degradation \cite{Dev15,Dev16,dMDS,MDS}. The paper \cite{HAR+} address the problem of small files. On the other hand, the standalone MDS cannot serve billions of client request and that is why most of the modern file system uses distributed MDS (dMDS) for better scalability. The most modern dMDS are Dr. Hadooop \cite{Dr}, CalvinFS \cite{CFS}, DROP \cite{Drop}, IndexFS and ShardFS \cite{IFS,SFS}, and CephFS \cite{SW04}. The issues of designing dMDS are Small file problem \cite{HAR+}, scalability \cite{Dr}, consistency, latency \cite{RAM,RAM1,LT11}, Hot-standby \cite{Dev16}, partition tolerance, network traffic, hotspot problem, and disaster recovery and management \cite{dMDS}. The table ~\ref{FS} shows the some important modern file system with respect to scalability, disaster recovery and management, hot-standby, single point of failure (SPoF), and types of metadata server (MDS).

\begin{table*}[ht]
\centering
\caption{Modern File System comparison, * denotes limited, $\times$ denotes no and $\checkmark$ denotes yes}
\scalebox{0.8}{
\begin{tabular}{p{4cm}p{1.5cm}p{1.5cm}p{1.5cm}p{1.5cm}p{1.5cm}p{2cm}}
\hline
\textbf{Name} & \textbf{Scalability} & \textbf{Disaster Recovery} & \textbf{Hot-standby} & \textbf{SPoF problem} & \textbf{POSIX-compliant} & \textbf{MDS} \\ \hline

HDFS \cite{HDFS} & * & $\times$ & $\times$ & $\checkmark$ & $\times$ & Standalone \\

GFS \cite{GFS} & * & $\times$ & $\times$ & $\checkmark$ & $\times$ & Standalone \\

CephFS \cite{SW04} & $\checkmark$ & $\times$ & $\checkmark$ & $\times$ & $\checkmark$ & Distributed\\ 

QFS \cite{Michael13} & * & $\times$ & $\times$ & $\checkmark$ & $\times$ & Standalone \\

BatchFS \cite{BatchFS} & $\checkmark$  & $\times$ & $\checkmark$ & $\times$ & $\checkmark$  & Distributed\\

Dr. Hadoop \cite{Dr} & $\checkmark$ & $\times$ & $\checkmark$ & $\times$ & $\times$ & Distributed \\

ShardFS \cite{SFS} & $\checkmark$ & $\times$ & $\checkmark$ &  & $\checkmark$ & Distributed \\

DMooseFS \cite{DMooseFS} & $\checkmark$ & $\times$ & $\checkmark$ & $\times$ & $\times$ & Distributed\\

CalvinFS \cite{CFS} & $\checkmark$ & $\checkmark$ & $\checkmark$  & $\times$ & $\checkmark$ & Distributed \\

GPFS \cite{GPFS} & $\checkmark$ & $\times$ & $\checkmark$ & $\checkmark$ & $\checkmark$ & Parallel\\

GlusterFS \cite{GlusterFS} & $\checkmark$ & $\checkmark$ & $\checkmark$ & $\times$ & $\checkmark$ & FUSE\\

DeltaFS \cite{DeltaFS} & $\checkmark$ & $\times$  & $\checkmark$ & $\times$  & $\times$ & LevelDB\\

\hline
\end{tabular}
}
\label{FS}
\end{table*}


\subsubsection{Object Storage for Big Data}
Object storage is a basic storage unit for applications which stores data as objects and as a logical collection of bytes on a storage device along with the methods for accessing and describing the characteristics of data. The object holds data and metadata \cite{Manek15}. The metadata is used to store the information about the content and context of the data. To access data, traditionally, the methods for input and output are specified explicitly in the application or use other external ways, and then the object storage system map these files into objects. The object is accessed using the globally unique identifier. We need to sacrifice the hierarchical layout of file and directory as a conventional file system has, since, there is no silver bullet. Object storage is designed to manage the heavy bit of unstructured files that need to be laid in. As well, it is best for archiving the information when it is not frequently accessed.

\subsubsection{Block Storage System}
The block storage system depends on the storage area network, and thus scaling is an issue this storage system. The block storage uses FCoE or iSCSI protocol to access the stored data and it is stored in the block of storage media. The block does not contain any information about the data, it contains only the raw data.

\subsubsection{Cloud Storage}
NIST defines Cloud computing as a model for enabling ubiquitous, convenient, on-demand network access to a shared pool of configurable computing resources that can be rapidly provisioned and released with minimal management effort or service provider interaction \cite{NIST}.  In the other word, Cloud computing is Internet computing that shares resources seamlessly. The cloud storage is a virtualized storage space where the actual data is stored in several servers. The cloud storage space can use object storage, file system, and hybrid storage system. The cloud storage provides high scalability, availability, security, fault tolerance, and cost-effective data services for those applications \cite{Huo11,Wu13}.
In that respect are three layers of Cloud Storage Architecture:
\begin{itemize}
\item The user application layer is the interface between the user and the virtual storage media.
\item  The Storage management layer is virtualization of the storage space. The virtualization manages the data, and create an illusion of simplicity and single storage space, while the storage space is rather complicated and span on several servers or geographical area.
\item The Storage resources layer deals with the actual data to store. Ordinarily, this layer uses file system or object storage system. The most advanced system uses a hybrid storage system which combines both storage systems. 
\end{itemize}

\subsection{Storage System Structure- NoSQL}
\label{NoSQL}

\begin{figure*}
\centering
\includegraphics[width=0.8\textwidth]{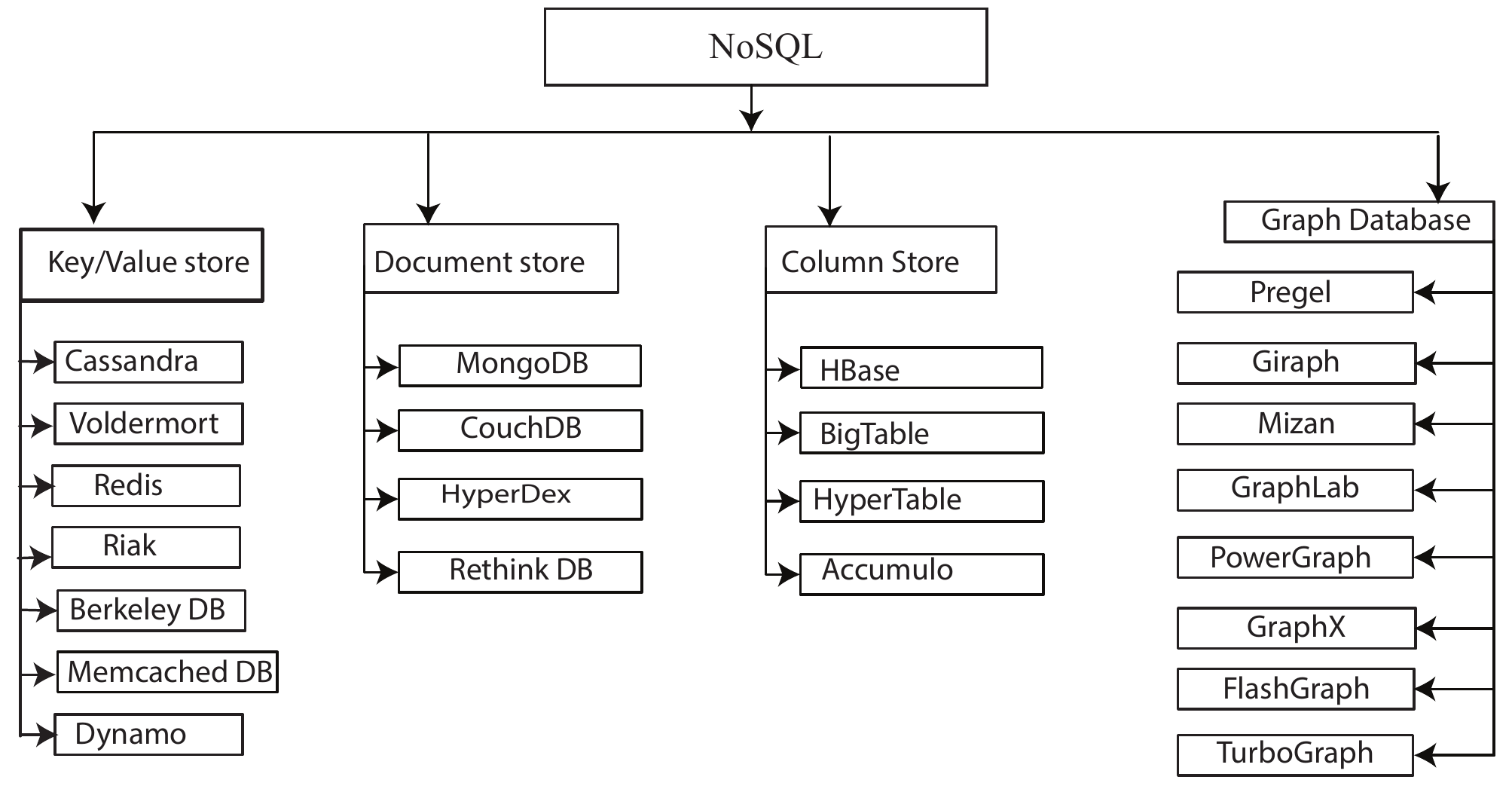}
\caption{NoSQL}
\end{figure*}

NoSQL \cite{Cattell10}, is Not Only SQL, emerging due to its urgent necessity in the industries. The NoSQL is the bigger dimension of SQL and non-SQL which is implemented for distributed or parallel computing. The alternative to RDBMS is NoSQL which provides high availability, scalability, fault-tolerance, and Reliability \cite{Lakshman10,Cassandra}. However, the NoSQL database does not perform well in OLTP due to ACID property requirement in OLTP process \cite{Storey}. The comparison of the NoSQL architecture is presented in the table ~\ref{tab3}.

\subsubsection{Key-value store}
The key-value store is the most elementary sort of storage where replication, versioning, and distributed locking are supported. The key-value store is very much useful in MapReduce environment.

\subsubsection{Column-Oriented Store}
The main issue in RDBMS is one column is exceeding RAM size, then the processing performance becomes very poor. Moreover, if it reaches to petabytes, then the conventional system does not work at all. To overcome this limitation, Google implements BigTable, which is scalable, flexible, reliable and fault-tolerance. In the current marketplace, there are a large amount of columnar database available, namely, HBase, HyperTable, Cassandra, Flink, eXtremeDB, and HPCC. This columnar family can ameliorate the processing speed in unstructured data as well as structured data.

\subsubsection{Document-oriented store}
The document store is similar to key-value store, except, the document store relies on the inner structure of the document to extract their metadata. XML database, for instance. The document store is a semi-structured database which provides fault-tolerance, and scalability in large-scale computation.

\subsubsection{Graph Database}
Graph database\cite{Kaliyar15} is more appropriate to deal with complex, densely connected, and semi-structured data. Graph databases are extremely helpful in the industries, for example, online business solution, healthcare, online media, financial, social network, communication, retail, etc. Graph database gives the response of complex queries in a few milliseconds because it stores the data in RAM. The shared-memory Big Graph processing engine is centralized whereas the other one is decentralized \cite{Arleo}. The decentralized graph processing engine is easy to scale. The issues and challenges of Big Graph are high-degree vertexes, sparseness, unstructured problems, in-memory challenge, poor locality, communication overhead, and load balancing \cite{DKS}.

\begin{table*}[ht]
\centering
\caption{Category comparison. Source \protect\cite{Cassandra}}
\begin{tabular}{p{2cm}p{2cm}p{2cm}p{2cm}p{2cm}p{2cm}}
\hline
\textbf{Name} & \textbf{Performance} & \textbf{Scalability} & \textbf{Flexibility} & \textbf{Complexity} & \textbf{Functionality}\\   \hline 
Key-value store & High & High & High & None & Variable\\ 
Column-store & High & High & Moderate & Low & Minimal\\ 
Row-store & High & High & Moderate & Low & Minimal\\ 
Document-store & High & High & Variable & Low & Variable\\ 
Graph Database & Variable & Variable & Low & High & Graph theory\\ 
RDBMS & Variable & Limited & Low & Moderate & Relational Model\\ \hline
\end{tabular}
\label{tab3}
\end{table*}

\subsection{Storage System Devices}

\begin{figure*}
\centering
\scalebox{0.8}{
\begin{tikzpicture}
    \begin{axis}[
    width=0.8\textwidth,
    height=8cm,
    ylabel=\$,
    ymajorgrids,
    axis y line*=left,
    ymode=log,
    log ticks with fixed point,
    x tick label style={rotate=45, anchor=east},
    xticklabels={,2008,,2009,,2010,,2011,,2012,,2013,,2014,,2015,,2016},
    xlabel={Year},
    domain=2008:2016,
    legend cell align=left,
        legend style={
                at={(0.35,1.05)},
                anchor=south east,
                column sep=1ex
        }
    ]
        \addplot[smooth,orange,mark=square] table[x=in,y=gdp] {costs.csv};\addlegendentry{RAM Cost per MB}
        \addplot[dashed,smooth,red,mark=triangle] table[x=in,y=gni] {costs.csv};\addlegendentry{HDD Cost per GB}
        \addplot[smooth,blue,mark=o] table[x=in,y=gns] {costs.csv};\addlegendentry{SSD cost per GB}
		
    \end{axis}
	\begin{axis}[
	width=0.8\textwidth,
    height=8cm,
    ylabel= Size level,
    axis y line*=right,
    log ticks with fixed point,
    x tick label style={rotate=45, anchor=east},
    xticklabels={,2008,,2009,,2010,,2011,,2012,,2013,,2014,,2015,,2016},
    legend cell align=left,
        legend style={
                at={(1,1.05)},
                anchor=south east,
                column sep=1ex
        }
    ]
    \addplot[smooth,purple,mark=x] table[x=in,y=ms] {costs.csv};\addlegendentry{RAM Size in GB}
    \addplot[smooth,green,mark=+] table[x=in,y=hs] {costs.csv};\addlegendentry{HDD Size in TB}
    \addplot[dotted,smooth,violet,mark=*] table[x=in,y=ss] {costs.csv};\addlegendentry{SSD size in TB}
	\end{axis} 	
\end{tikzpicture}
}
\caption{RAM, SSD, and HDD cost per MB. Source \protect\cite{MPS,MPS1,MPS3,MPS2} respectively}
\label{cost}
\end{figure*}
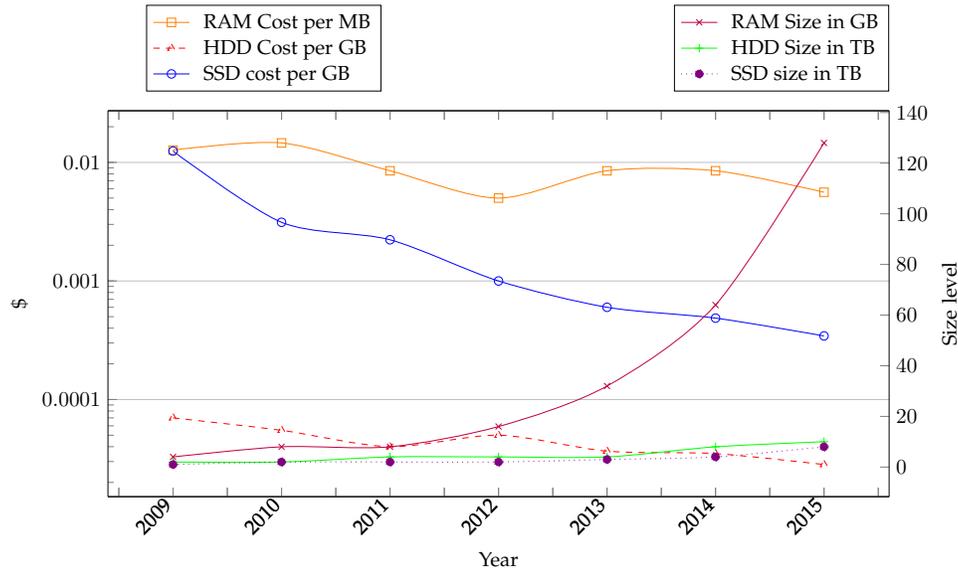

It is the time to do research on latency and it is the hot potato in the research field. The RAMCloud \cite{Ousterhout15}, Memcached, and Spark \cite{Spark}  deal with the latency issues. The performance of a system depends on latency and the latency is the great impact factor of a performance of a particular system. The latency of RAM is lower than the SSD and HDD. The most prominent field of research is latency to reduce, and eventually, SSD has turned up. Even though, the SSD cannot be as fast as RAM, but still reducing the latency. The research challenge of HPC requires the highest performance with the lowest cost in \$. In-memory system requires SSD or HDD support for durability, otherwise, system shutdown causes data lost and it is not tolerated at any cost. The race is now among Cache memory, RAM and SSD/PCIe PCM. The main component of performance is RAM, but unfortunately, cost of RAM does not decrease satisfactory as in SSD or HDD as shown in the figure \ref{cost}.

As the figure ~\ref{cost} shown, the RAM cost is falling very slowly, and its size is increasing exponentially. On the other hand, The HDD and SSD cost is really low and continuously falling. Furthermore, the size of SSD and HDD is slowly rising.

\subsubsection{Cache Memory}
No doubt, cache memory is the fastest memory devices. The challenges of designing algorithms for distributed system are to increase the cache hit ratio. There are numerous researcher working on cache-aware algorithm, namely, scheduling, MDS designing.

\subsubsection{Primary Memory}
Designing in-memory Big Data system is an open challenge. There are many such systems has been unleashed, such like, RamCloud, H-Store, Big Graph Analytic software.

\subsubsection{SSD}
The Solid-State Device (SSD) is the most advanced storage device for Big Data technology. The SSD is the hybrid version of RAM and secondary memory. The designing of the system must utilize recent the technology, such that the outcome should gain highest possible benefit from them. The design decision must ensure that the new recent technology should not degrade the performance of the system.

\subsubsection{HDD}
The most common form of storage device is Hard Disk (HDD) now-a-days. The HDD has a latency problem due to finding the track and sector.

\subsubsection{Tape}
The tape is the most inexpensive and bulk storage device, but the read/write performance is really inadequate. This type of storage device has significant advantage in implementing a storage system for backup purpose.

\section{Big Data Management}

\begin{figure}[ht]
\centering
\includegraphics[width=0.45\textwidth]{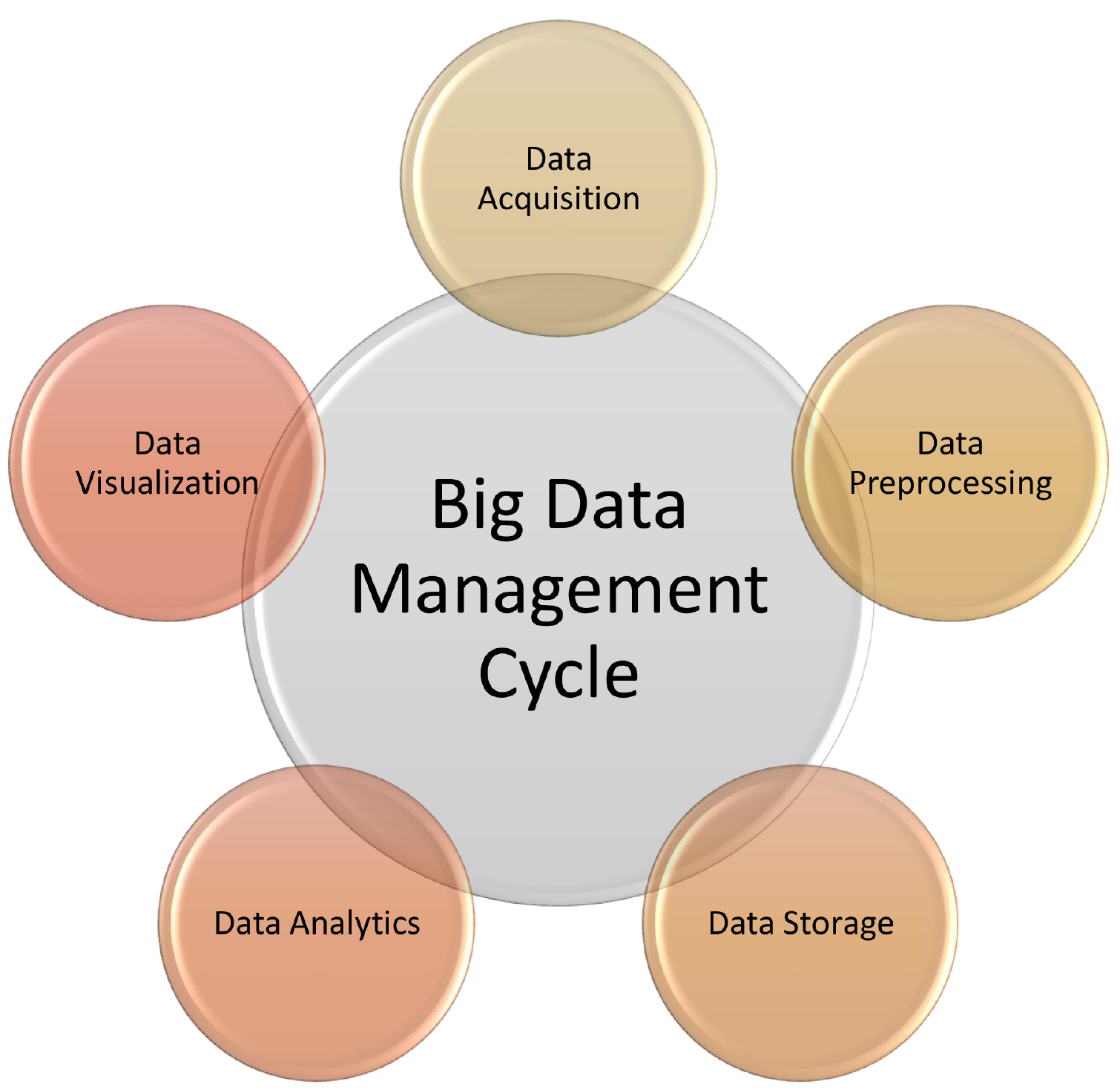}
\caption{Big Data management cycle}
\end{figure}

\subsection{Data Acquisition}
Data acquisition \cite{Acquisition} is the process of collecting, filtering and removing any noise from data before they can be stored in any data warehouse or any storage system. It adopts adaptive and time efficient algorithms for processing of high value data. For data acquisition, frameworks are available that are based on predefined protocols. However, many organizations that depend on big data processing have developed their own enterprise-specific protocols. Most commonly used open protocol is Advanced Message Queuing Protocol (AMQP). It satisfies a series of requirements compiled by 23 companies. And it became an OASIS standard in October 2012. It has characteristics such as ubiquity, safety, fidelity, applicability, interoperability, and manageability. In addition, lots of software tools are also available for data acquisition (e.g. Storm, S4).

\subsection{Data Preprocessing}
Data preprocessing \cite{preprocessing} is the set of techniques used before the application of data processing techniques. It removes data redundancies and inconsistencies, and make it suitable for application of data processing algorithms. Some data preprocessing approaches are Dimensionality reduction and Instance reduction. Dimensionality of data refers to the instances of the data. And Dimensionality reduction include Feature selection and Space transformations. Whereas, Instance reduction refers to reduction of size of data set. It include Instance selection and Instance generation. Additionaly, In Big Data, MLlib \cite{MLlib} is used for data preprocessing in Spark. MLlib is a powerful machine learning library that helps in use of Spark in data analysis.

\subsection{Data Storage}

\begin{table}[ht]
\caption{List of shared and non-shared storage system}
\centering
\resizebox{0.75\textwidth}{!}{
\begin{minipage}{\textwidth}
\begin{tabular}{p{3cm}p{8cm}}
\hline
\textbf{Name} & \textbf{Name of File Systems} \\ \hline \hline
Shared-Nothing Architecture & QFS, HDFS, GFS, GPFS, BeeGFS, Ceph, GlusterFS, OneFS, OrangeFS, MooseFS, ObjectiveFS, PanFS, Parallel Virtual File System, Windows Distributed File System (DFS), and XtreemFS  \\ \\
Shared-Everything Architecture & Lustre, PVFS, GPFS, VxCFS, Quick File System, VMFS, BWFS, GFS2, and OCFS2 \\ \hline

\end{tabular}
\end{minipage}
}
\label{tab8}
\end{table}

\subsubsection{Shared-nothing Architecture}	
The shared-nothing architecture does not share its resources to others. The resources are HDD, SSD, and RAM. The significant advantage of shared-nothing is fine-grained fault tolerance, scalability, and maximize the parallelism. The table ~\ref{tab8} shows the technology that uses Shared-nothing architecture and shared-everything architecture.

\subsubsection{Shared-everything Architecture}
Sharing always cause synchronization, whatever the implementation is.
\begin{itemize}
	\item Shared Memory- Even though the fastest inter-process communication using shared memory, but there arises some synchronization issues. Careful design can lead to the best performance of the resources.
    \item Shared Disc- The sharing storage devices implemented very widely.
    \item Shared Both- The hybrid system always exists.
\end{itemize}

\subsection{Data Analytics}
The Big Data Analytics is an logical analysis on large set of data for specific purposes. It requires Data Mining algorithms which is classified into four key categories, namely, Machine Learning, Statistic, Artificial Intelligence, and data warehouse. The Big Data analytics requires Statistics and Machine Learning. The Machine Learning (ML) algorithms are applied to analyze and predict on very large dataset. Moreover, the ML algorithms take huge time in execution. The ML algorithms require new technologies to execute a mammoth sized data. The Big Data tools efficiently handle ML algorithms in very large scale data-intensive computation.

\begin{figure}
\centering
\scalebox{0.8}{
\newcolumntype{C}[1]{>{\centering}p{#1}}
\begin{forest}
 for tree={
  		if level=0{align=center}{
    		align={@{}C{30mm}@{}},
  		},
  		grow=east,
  		draw,
  		font=\sffamily\bfseries,
  		edge path={
    		\noexpand\path [draw, \forestoption{edge}] (!u.parent anchor) -- +(3mm,0) |- (.child anchor)\forestoption{edge label};
  		},
  		parent anchor=east,
  		child anchor=west,
  		l sep=10mm,
	}
  [Big Data \\Analytics
  	[User\\Behavioral]
    [Diagnostic
    	[Detection]
        [Prevention]
    ]
    [Prescriptive]
    [Decision Analytics
    	[Decision \\possibility]
        [Decision \\Assistance]
    ]   
    [Predictive
    	[Possibility]
    	[Forecasting]
    ]
  	[Descriptive
        [Discovery]
        [Real-time]
    ]     
  ]
\end{forest}
}
\caption{Taxonomy of Big Data Analytics}
\label{taxanalytics}
\end{figure}
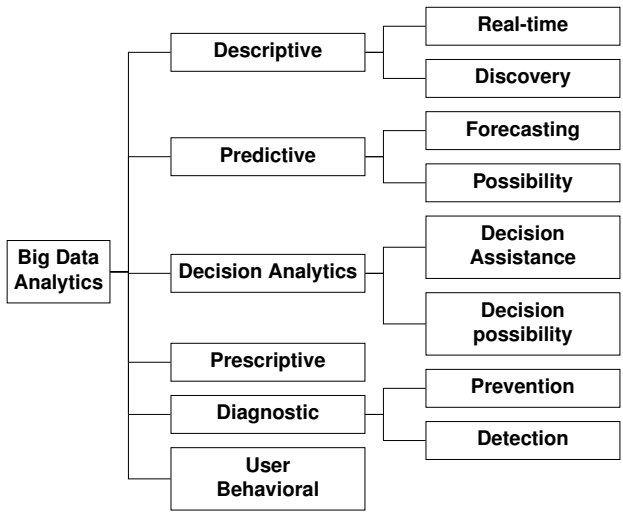

\subsection{Data Visualization}
\begin{figure*}[ht]
\centering
\includegraphics[width=0.8\textwidth]{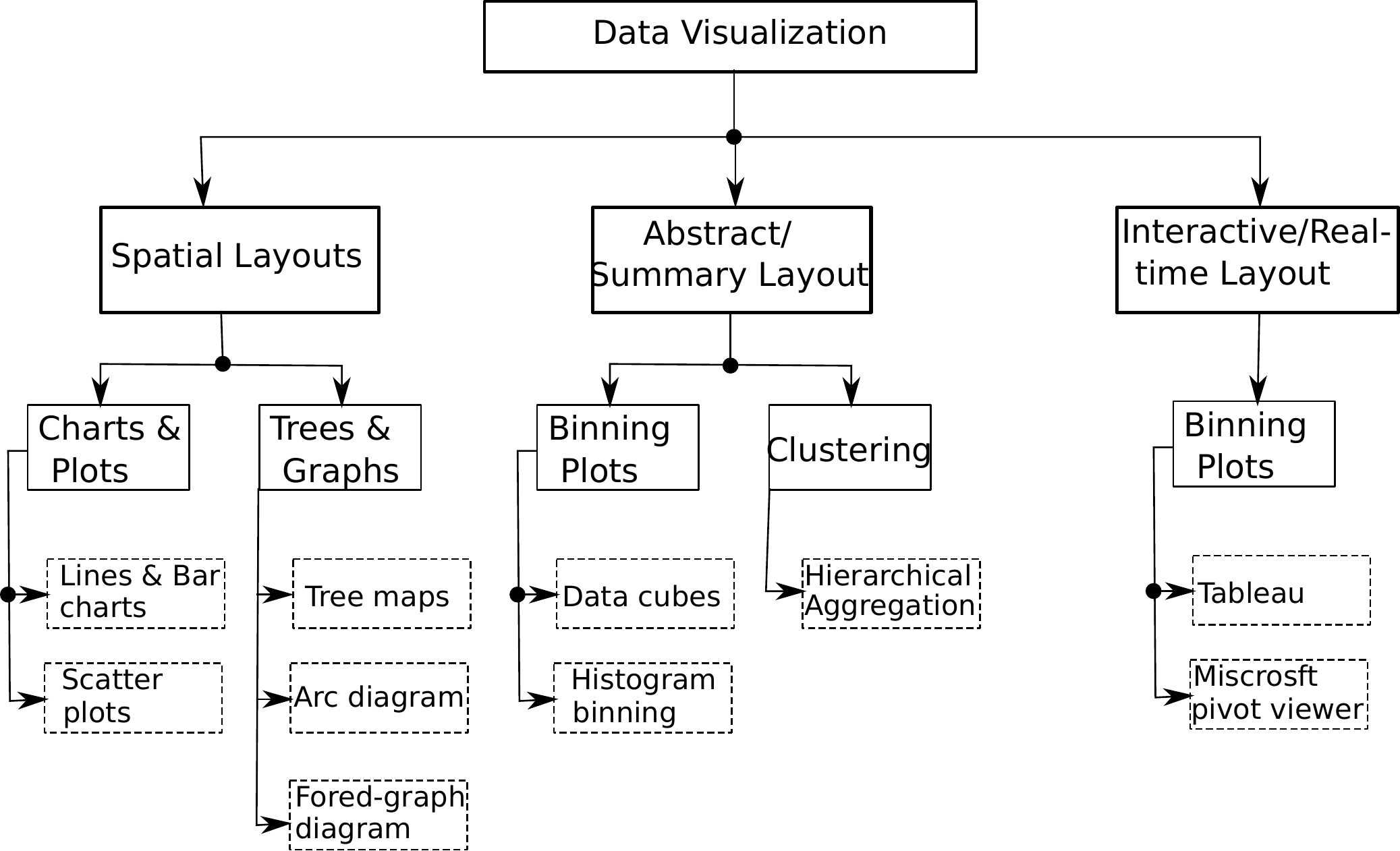}
\caption{Taxonomy of Data Visualization. Source \cite{CSA}}
\end{figure*}
Spatial Layout visualization techniques refers to formulas that maps an input data uniquely to a specific point on the coordinate space. Popular visualization techniques under it can be classified into chart and plots, and trees and graphs. Example of some techniques coming under chart and plots are line and bar chart, and scatter plots. Again latter have tree maps, arc diagram, and forced graph drawing.

Abstract/Summary Visualization techniques does abstraction or summarization before representation of data. For that scaling is one of the technique. It is done for easy understanding of data (e.g. 1cm=1km in map) which helps in finding meaningful correlations among them. Common technique for data abstraction is binning it into histograms or data cubes. Its advantage is providing a compressed, reduced dimensional representation of data. Likewise, its techniques can also be classified into another group, clustering (e.g Hierarchical Aggregation).

Interactive/Real-Time Visualization techniques are recent techniques that have ability to adapt to user interactions in real-time. Such techniques have capability to take less than a second for a real-time navigation of data by a user. Such techniques are powerful because they can quickly discover important details in the data that helps to verify different data science theories. For example Microsoft Pivot Table and Tableau, enables to pivot the data in Microsoft Excel, text file, .pdf, and Google Sheets data sources from crosstab format into columnar format for easy analysis.

\section{Data Mining and Machine Learning}
\subsection{Big Data Mining}

\begin{figure}[ht]
\centering
\includegraphics[width=0.45\textwidth]{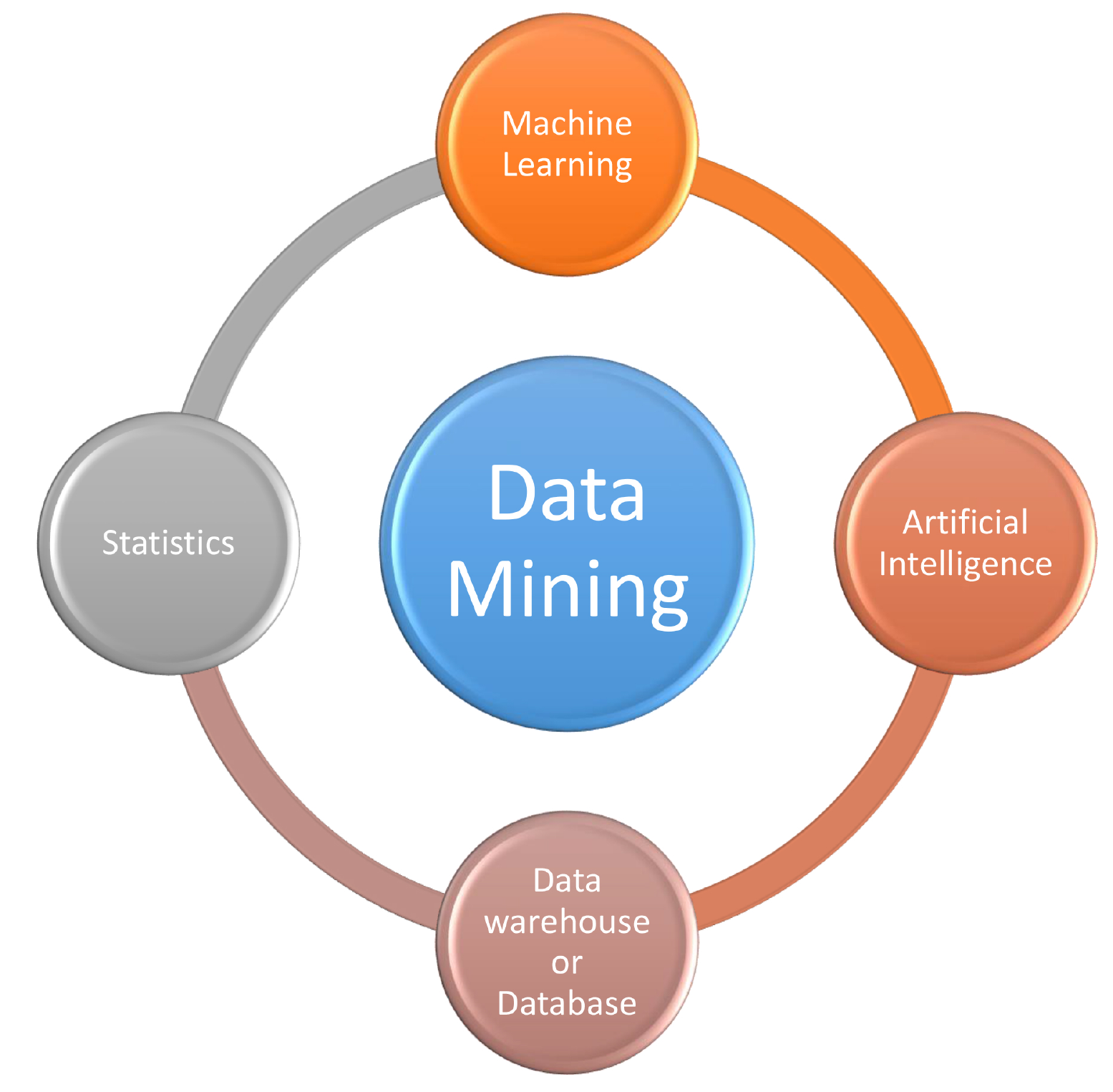}
\caption{Taxonomy of Data Mining}
\end{figure}

Too big data, too frequent data incoming, too frequent data changes, and too complex data. These are the data to be envisioned. Visualization of few MB data is not a big deal, but the huge data is a big sight. The jewels have to be mined from the massive amount of data, analyzed for business purposes- such like business prediction, and use for growth of business.

Big data is the most hyped word in recent times which depicts prodigious volume of data consisting of various data formats and it is very difficult to process them using conventional mode of database, software methodologies and also it is successful enough to attract the attention of technology  dwellers whereas data mining is the method or technique to mine useful information in the form of patterns that enlighten our vision about data and its utilities in every sphere of life starting from life science, business, etc.  Although big data and data mining are the two different perspectives of modern technology but they have a relationship of dealing with massive quantity of data, for example, Twitter termed their data mining experience as Big Data Mining which was treated as data mining \cite{JLin}.

Simply storing huge volume of data from time to time without using it for organizational benefit is merely wastage of resources. So we should use this data for acquiring knowledge which can be useful for future work. But it is very difficult to handle such enormous volume of data and analyze it, under such scenario comes the concept of data mining for getting specific information from the big data. Many algorithms and techniques have been applied to mine useful information from the deep ocean of data. Big data is an ocean of enormous volume of data but mining precious and useful information out of this can be done with the help of efficient use of data mining techniques such as classification, clustering, outlier detection, association rule, etc. We encounter different levels of difficulty in mining useful information from large data sets as we need to handle our data with the changing requirement of technology. But we get our required set of analyzed results with advent use of data mining techniques, for instance, sentimental analysis \cite{Salehan}.

The data mining techniques applied on the Big data should be applicable for any kind of data. The various data mining algorithms such as C4.5, K-Means, Support Vector Mechanism (SVM), Apriori, KNN, CART, Naive Bayes, Page Rank, etc. are widely applied in the field of Big Data Analytics. Likewise, the OLAP over Big Data is evolving \cite{Cuzzocrea13}.

\subsection{Machine Learning}
\begin{figure*}
\centering
\includegraphics[width=\textwidth]{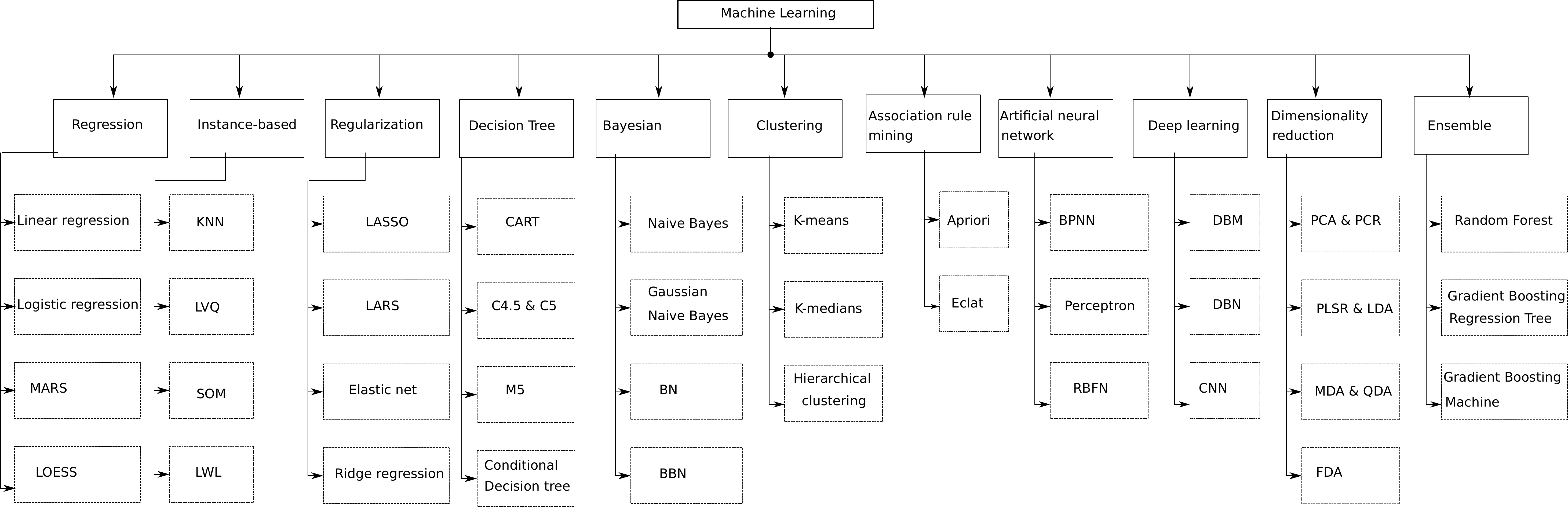}
\caption{Taxonomy of Machine Learning Algorithms. Source\cite{Jason}}
\label{mla}
\end{figure*}

Machine learning is a field of computer science that try to enable computers with the ability to learn without explicitly programming them. It$’$s algorithms can be classified as following (Fig.14). It is categorized into eleven categories depending on the characteristics of the algorithm as shown in the figure \ref{mla}, albeit, the ML algorithms are categorized in three key categories, namely, supervised, unsupervised and semi-supervised ML algorithm. The Big Data Analytics or Big Data Security analytics fully depends on Data Mining, where the Machine Learning techniques are subset of Data Mining. The supervised learning algorithms are trained with a complete set of data and thus, the supervised learning algorithms are used to predict/forecast. The unsupervised learning algorithms starts learning from scratch, and therefore, the unsupervised learning algorithms are used for clustering. However, the semi-supervised learning combines both supervised and unsupervised learning algorithms. The semi-supervised algorithms are trained, and the algorithms also include non-trained learning.

\begin{itemize}
\item \textbf{Regression algorithm:} It predict output values using input features of the data provided to the system. Most popular algorithms under it are Linear Regression, Logistic Regression Multivariate Adaptive Regression Splines (MARS) and Locally Estimated Scatter plot Smoothing (LOESS).

\item \textbf{Instance based algorithm:} It compares new problem instances with instances in training, that are stored in memory. Common algorithms can be k-Nearest Neighbor (kNN), Learning Vector Quantization (LVQ), Self-Organizing Map (SOM) and Locally Weighted Learning (LWL).
 
\item \textbf{Regularization algorithm:} It is a process of providing additional information to prevent overfitting or solve an ill-posed problem. Algorithms which are common under it are Least Absolute Shrinkage and Selection Operator (LASSO), Least-Angle Regression (LARS), Elastic Net and Ridge Regression.

\item \textbf{Decision tree algorithm:} It solves the problem using tree representation. Some popular algorithms are Classification and Regression Tree (CART), C4.5 and C5.0, M5, and Conditional Decision Trees.

\item \textbf{Bayesian algorithm:} It is based on Bayesian method. Popular algorithms are Naive Bayes, Gaussian Naive Bayes, Bayesian Network (BN), and Bayesian Belief Network (BBN).

\item \textbf{Clustering algorithm:} Grouping of data points based on similar features. Some popular algorithms are K-Means, K-Medians, and Hierarchical Clustering.

\item \textbf{Association Rule Learning Algorithms:} It is used to discover relationship between data points. Some common algorithms under it are Apriori algorithm, and Eclat algorithm.

\item \textbf{Artificial Neural Network Algorithms:} It is based on working of biological neural networks. Example of some popular algorithms are Back-Propagation Neural Network(BPNN), Perceptron, and Radial Basis Function Network (RBFN).

\item \textbf{Deep Learning Algorithms:} It uses unsupervised learning to set each level of hierarchy of features using features discovered at previous level. It has some popular algorithms such as Deep Boltzmann Machine (DBM), Deep Belief Networks (DBN), and Convolution Neural Network (CNN).

\item \textbf{Dimensionality Reduction Algorithms:} They reduces the number of feature by obtaining a set of principal variables. Some algorithms commonly under it are Principal Component Analysis (PCA) and Principal Component Regression (PCR), Partial Least Squares Regression (PLSR) and Linear Discriminant Analysis (LDA), Mixture Discriminant Analysis (MDA) and Quadratic Discriminant Analysis (QDA) and Flexible Discriminant Analysis (FDA).

\item \textbf{Ensemble Algorithms:} It combines multiple learning algorithms to obtain better predictive performance. Some well know algorithms are Random Forest, Gradient Boosted Regression Trees (GBRT), and Gradient Boosting Machines (GBM).

\end{itemize}

\section{Security and Privacy}
Analysis of Big data provides a large volume of knowledge which can be used many ways for the betterment of individual, society, nation, and even the world. Nowadays people have become open with their thoughts to the world. So it is the responsibility of the people who are using these data, to protect the data and prevent others from misusing it. 
\begin{figure*}
\centering
\includegraphics[width=0.9\textwidth]{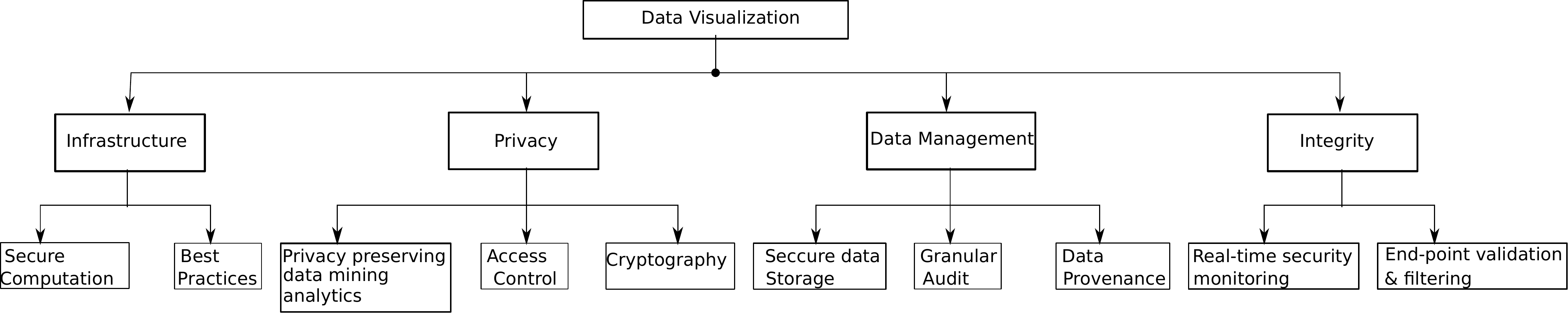}
\caption{Security and privacy. Source \cite{CSA}}
\end{figure*}

The Security and Privacy challenges for big data can be classified into four groups (Fig.15) such as Infrastructure Security, Data Privacy, Data Management, and Integrity and Reactive Security. 
\subsection{Infrastructure Security}
Infrastructure Security\cite{security} in big data systems includes securing distributed computations and non-relational data stores. As Hadoop framework is most commonly used for distributed system so its security is mostly researched to make it robust to threats. Example of such a security model is G-Hadoop that implements users’ authentication and some security mechanisms in a simplified way to protect the system from traditional attacks. Additionally, Distributed system uses parallelism for computation and storage of high volume of data. Common example is Mapreduce framework so its security at mapper and reducer phase is important. For that two main attack preventive measures are securing the mappers and securing the data during the presence of untrusted mapper. Also big data use non-relational data stores so focus on its security is also important. For such data stores NoSQL is used which don't have security provision so security in middleware is used. However using clustering in NoSQL impose some security threats to it.

\subsection{Data Privacy}
Organisations want to protect the privacy of data and also want to make a profit out of it. With that aim several techniques and mechanisms were developed such as Privacy preserving data mining and analytics, Cryptography, Access Control. Now, Privacy Preserving data mining and analytics means efficiently finding valuable data which are prone to misuse. And Cryptography is the most commonly used mechanism for data security. Some examples are Homomorphic Encryption (HE), secure Multi-Party Computation (MPC), and Verifiable Computation (VC). Whereas Access control is to stop undesirable users from accessing the system.  

\subsection{Data Management}
Data Management\cite{BigD} comes into picture after data is stored in big data environment. From security point of view it include Secure data storage and transaction logs, Granular audits, and Data provenance. Security to data storage and transaction logs are necessary. As multi-tiered storage media are used to store data and transaction logs. So, Auto-tiering is used to move data in data storage because of huge size of data. But this disable the system to keep track of where data is stored. So new mechanisms were developed for security and round the clock availability of data. Similarly transaction logs security are essential as they contain all the changes made in system. Similarly, Granular audits are crucial as it provide information related to attacks. It have information about what happened and what preventive measures can be taken. It also can be used for agreement, regulation and forensic investigation. Likewise, Data provenance provide information about data and its origin that include input data, entities, systems and processes influencing that particular data. But its complexity increases as provenance enabled programming environments in Big Data applications produces complex provenance graphs. Analysis of such complex information is difficulty but required for security purposes.

\subsection{Integrity}
Big data collects data from various sources and stores them in various formats. So there comes the importance of integrity of data which is the accuracy, consistency, and trustworthiness of data. As shown in the figure integrity can be classified into Endpoint validation and filtering, and Real-time security monitoring. Before the data processing it is essential to validate the authenticity of input data otherwise system may be processing bad data. For this purpose Endpoint validation and filtering is essential for producing good and trustworthy results. Real-time security monitoring is used to monitor big data infrastructure. However the security devices generates lots of security alarms and false positives. And as it is big data these may increases further more in volume \cite{integrity}.

\section{Conclusion}
The Big Data is game changer paradigm in data-intensive field and it is very wide area to study. The Big Data is a data silos. It is very difficult to process, manage and store the data silos. The data silos are formed not only in core computing area, but also science, engineering, economy, government, environment, society, etc. There are a variety of processing engine to process mammoth sized data efficiently and effectively. These processing engines are developed based on their requirements and characteristics. Moreover, different kind of storage engines are also emerging, for instance, file system and object storage system. Besides, machine learning algorithms are integrated with Big Data. As a consequence, the Big Data Analytics is born.
\balance
\bibliographystyle{IEEEtran}
\bibliography{mybibfile}

\end{document}